\documentclass[12pt]{article}

\input epsf.tex
\newcommand{\be}{\begin{equation}}
\newcommand{\ee}{\end{equation}}
\newcommand{\bea}{\begin{eqnarray}}
\newcommand{\eea}{\end{eqnarray}}

\begin{document}

\bigskip 
\begin{titlepage}

\begin{flushright}
UUITP-12/02\\ 
hep-th/0210058
\end{flushright}

\vspace{1cm}

\begin{center}
{\Large\bf On the consistency of de Sitter vacua\\}

\end{center}
\vspace{3mm}

\begin{center}

{\large
Ulf H.\ Danielsson} \\

\vspace{5mm}

Institutionen f\"or Teoretisk Fysik, Box 803, SE-751 08
Uppsala, Sweden

\vspace{5mm}

{\tt
ulf@teorfys.uu.se \\
}

\end{center}
\vspace{5mm}

\begin{center}
{\large \bf Abstract}
\end{center}
\noindent
In this paper the consistency of the de Sitter invariant $\alpha $-vacua, which have been introduced as simple tools to study the effects
of transplanckian physics, is investigated. In particular possible non renormalization problems
are discussed, as well as non standard properties of Greens functions. We also discuss the non thermal properties
of the  $\alpha $-vacua and the necessity of $\alpha$ to change. The conclusion is that 
non of these problems necessarily exclude an application of the $\alpha $-vacua to inflation.


\vfill
\begin{flushleft}
October 2002
\end{flushleft}
\end{titlepage}
\newpage


\section{Introduction}

\bigskip

The main driving force behind recent work on transplanckian physics in
cosmology, is to find out whether physics beyond the Planck scale can give
effects on the CMBR spectrum, [1-30]. A crucial ingredient in inflationary
cosmology is that microscopic quantum fluctuations are magnified by
inflation into macroscopic seeds for galaxy formation. In standard inflation
the modes of the inflaton field can be carried back in time to eras when
they were much smaller than the Planck scale. For small scales the expansion
of the universe can be ignored and a unique vacuum can be chosen for the
inflaton. This is the Bunch-Davies vacuum. But the construction ignores the
Planck scale and the natural expectation that physics beyond the Planck
scale is very different from physics at low energies, and not possible to
describe using a quantum field theory. The hope would be that transplanckian
physics modifies the standard scenario and that the effects are magnified
through inflation to leave visible imprints in the CMBR fluctuations.

There are two main approaches to investigate whether the above proposal is
viable. One possibility is to use specific models of transplanckian physics,
examples include models involving non commutative geometry, and investigate
how this will change the predictions of inflation. Another possibility is to
leave the details of the transplanckian physics for later work, and impose a
cutoff on the theory at the Planck scale, or some other scale like the
strings scale, where fundamentally new physics is expected. Our ignorance of
the high energy physics is then encoded in the choice of initial conditions
for the field modes when they start out at planckian size. Contrary to the
standard scenario, the initial conditions are imposed in a situation where
the time dependence of the background can not be ignored. Various natural
ways of choosing the vacuum, minimal uncertainty, adiabatic to all orders
etc., now give different results. The Bunch-Davies vacuum remains as a
possibility but there are also other possibilities that also can be argued
to be natural. A conservative approach is then to investigate the span of
possibilities and see what effects, if any, they give on the CMBR. In \cite
{Danielsson:2002kx} it was argued that the generic size of the effects is
expected to be of order $H/\Lambda $ in the spectrum compared with the
Bunch-Davies vacuum. To claim smaller effects one would need specific
information about the nature of transplanckian physics. It is important to
note that the generic effect is not a simple change in normalization of the
spectrum but has a definite signature in the form of a modulation.

The simplified approach discussed in \cite{Danielsson:2002kx}\cite
{Danielsson:2002qh}, which is not tied to specific models of transplanckian
physics, allows for a detailed examination of the consequences and viability
of the transplanckian proposal. As observed in \cite{Danielsson:2002qh} it
essentially amounts to an investigation of the physics of the de Sitter
invariant vacua introduced in \cite{chernikov}\cite{Mottola:ar}\cite
{Allen:ux}\cite{Floreanini:1986tq}, some times called the $\alpha $-vacua.
These vacua have also recently been discussed in the context of de Sitter
holography, \cite{Bousso:2001mw}\cite{Spradlin:2001nb}.

The purpose of this note is to address recent discussions in the literature,
in particular \cite{Banks:2002nv}\cite{Einhorn:2002nu}\cite{Kaloper:2002cs},
where the consistency of the $\alpha $-vacua is questioned. In the first two
of the above mentioned works it has been pointed out that quantum field
theory with $\alpha $-vacua is not well understood. In particular there are
problems with renormalizability and the definition of loop amplitudes, and
there are also peculiarities in certain Greens functions that are claimed to
suggest that the vacua does not make physical sense.

The outline of the note is as follows. In section two we review the
construction of the $\alpha $-vacua, in section three we focus on the
problems of non renormalizability, in section four we discuss the large
scale structure of the Greens functions, in section five how $\alpha $ might
change with time, in section six the non-thermal nature of the $\alpha $%
-vacua and, finally, we end with some conclusions.

\section{Review of $\protect\alpha $-vacua}

\bigskip

Let us briefly review how the $\alpha $-vacua are constructed. We will focus
on a single inflaton field, $\phi \left( x\right) $, and its quantization.
The Wightman function $G^{+}\left( x,x^{\prime }\right) $ for the inflaton
field is defined as 
\begin{equation}
G^{+}\left( x,x^{\prime }\right) =\left\langle \Omega \right| \widehat{\phi }%
\left( x\right) \widehat{\phi }\left( x^{\prime }\right) \left| \Omega
\right\rangle =\int d^{3}k\phi _{\mathbf{k}}\left( x\right) \phi _{\mathbf{k}%
}^{\ast }\left( x^{\prime }\right) ,  \label{wightman}
\end{equation}
where the field is expanded in modes as 
\begin{equation}
\widehat{\phi }\left( x\right) =\int d^{3}k\left[ \phi _{\mathbf{k}}\left(
x\right) \widehat{a}_{\mathbf{k}}+\phi _{-\mathbf{k}}^{\dagger }\left(
x\right) \widehat{a}_{\mathbf{k}}^{\dagger }\right] .  \label{phiexp}
\end{equation}
Operators are equipped with hats in all expressions. We then assume that\
these modes are obtained through Bogolubov transformations from the
Bunch-Davies modes according to 
\begin{equation}
\phi _{\mathbf{k}}\left( x\right) =A\phi _{\mathbf{k,}BD}\left( x\right)
+B\phi _{-\mathbf{k,}BD}^{\dagger }\left( x\right) ,
\end{equation}
with 
\begin{equation}
\left| A\right| ^{2}-\left| B\right| ^{2}=1.
\end{equation}
A convenient parametrization is to write 
\begin{equation}
A=\frac{1}{\sqrt{1-e^{\alpha +\alpha ^{\ast }}}}\quad B=\frac{e^{\alpha }}{%
\sqrt{1-e^{\alpha +\alpha ^{\ast }}}},
\end{equation}
where $\alpha =-\infty $ is the Bunch-Davies vacuum. For convenience we
write 
\begin{equation}
\phi _{\mathbf{k,}BD}\left( x\right) =\phi _{k,BD}\left( \eta \right) e^{i%
\mathbf{k}\cdot \mathbf{x}},
\end{equation}
and demand, for simplicity, that 
\begin{equation}
\phi _{k,BD}\left( \eta \right) =\phi _{k,BD}^{\dagger }\left( -\eta \right)
.
\end{equation}

To proceed, let us recall a few elementary aspects of de Sitter geometry.
The metric in terms of coordinates useful for inflationary cosmology is
given by 
\begin{equation}
ds^{2}=dt^{2}-a\left( t\right) ^{2}d\mathbf{x}^{2},  \label{metric}
\end{equation}
where $\ a\left( t\right) =e^{Ht}$ is the scale factor. In terms of the
conformal time $\eta =-\frac{1}{aH}$ the metric becomes 
\begin{equation}
ds^{2}=\frac{1}{H^{2}\eta ^{2}}\left( d\eta ^{2}-d\mathbf{x}^{2}\right) .
\label{confmet}
\end{equation}
The inflationary universe can be viewed as the upper right triangle of the
full de Sitter space, which has a Penrose diagram with the shape of a
square, where $\eta \rightarrow -\infty $ corresponds to the Big Bang, while 
$\eta \rightarrow 0$ is the infinite future. The lower half of de Sitter
space can be covered if we also consider positive $\eta $. A crucial
ingredient in the study of the $\alpha $-vacua is the antipodal map, which
in the above coordinates takes $x=\left( \eta ,\mathbf{x}\right) $ into $%
\overline{x}=\left( -\eta ,\mathbf{x}\right) $, and acts like a map between
the two halves of de Sitter space.

With the help of the antipodal map the Bogolubov transformation can now be
written 
\begin{equation}
\phi _{\mathbf{k}}\left( x\right) =A\phi _{\mathbf{k,}BD}\left( x\right)
+B\phi _{\mathbf{k,}BD}\left( \overline{x}\right) ,
\end{equation}
and the Wightman function becomes

\begin{equation}
G^{+}\left( x,x^{\prime }\right) =\left| A\right| ^{2}G_{BD}^{+}\left(
x,x^{\prime }\right) +\left| B\right| ^{2}G_{BD}^{+}\left( x^{\prime
},x\right) +AB^{\ast }G_{BD}^{+}\left( x,\overline{x}^{\prime }\right)
+BA^{\ast }G_{BD}^{+}\left( \overline{x},x^{\prime }\right) .  \label{planw}
\end{equation}

After this brief review of the $\alpha $-vacua we will now turn to the main
subject of the paper: do the $\alpha $-vacua make physical sense?

\section{Are loop amplitudes ill defined?}

\bigskip

In \cite{Banks:2002nv}\cite{Einhorn:2002nu} problems with the definition of
loop amplitudes were pointed out. In both papers one loop amplitudes were
investigated and failed to give finite and well defined results. In \cite
{Banks:2002nv} it was shown that non local counter terms in the action,
involving insertions at image points, were needed. In \cite{Einhorn:2002nu}
the analytic structure of the Greens functions indicated that pinched
singularities in the loop integrations make the results ill defined. These
problems are clearly of great interest and it is a challenge to make sense
out of the field theory under these circumstances. But none of these
problems are necessarily relevant to the issue of transplanckian physics in
cosmology.

The reason is simple. The whole point with the transplanckian physics, as
explained in the introduction, is to see whether effects beyond quantum
field theory can be relevant for the detailed structure of the fluctuation
spectrum of the CMBR. Without a planckian cutoff there is no reason to
impose initial conditions at any finite scale. The only natural procedure is
to go to the infinite past, when the modes are infinitely small, and make
the choice there. For these small scales the expansion of the universe is
irrelevant and there is a unique natural choice of vacuum. This would be the
end of the story in a world without a Planck scale (and dynamical gravity)
in which quantum field theory could be trusted to all energies. The issue of 
$\alpha $-vacua, and transplanckian physics, would never arise. In the real
world we do expect quantum field theory to break down at high enough energy
to be replaced by something else, presumably string theory. To find out what
kind of effects this new physics might have, we can try to modify the
transplanckian physics by hand. One way to gain control over the situation
is to impose a cutoff and assume quantum field theory to hold for energies
below the cutoff. It is now the issue of vacuum choice becomes important.
Since\ the Planck scale is not infinitely smaller than the inflationary
Hubble scale, the time dependence of the background has to be taken into
account, and there is no unique natural vacuum. In fact, there are certain $%
\alpha $-vacua that are as natural a product of the unknown transplanckian
physics as the Bunch-Davies vacuum. The modest proposal behind \cite
{Danielsson:2002kx} is simply that we should allow for this uncertainty in
the possible outcome and investigate the consequences.

To summarize, the issue of $\alpha $-vacua only appears when one takes a
planckian cutoff into account. This is perfectly consistent with the results
of \cite{Einhorn:2002nu}. \textit{The }$\alpha $\textit{-vacua not only
needs a planckian cutoff to be of physical relevance, they need a planckian
cutoff to make physical sense. }In other words, the quantum field theory
loop amplitudes needs planckian input to give well defined answers.

In this context one should note that a rough cutoff in the sum over momenta
and energies in (\ref{wightman}) at the Planck scale, regulates the standard
singularities in the Greens functions that occur when the insertions can be
joined by a light ray. But it is not really these singularities that make
trouble, it is rather those that are due to the image charges. But, as is
easily seen, all terms in the Wightman function in (\ref{planw}) are
regulated with this procedure. With these regulated Greens functions finite
loop amplitudes can be constructed, even though their values are sensitive
to exactly how the cutoff is implemented, that is, the results depend on
transplanckian physics. The main point is that in a quantum gravity theory,
where the Planck scale plays an important role, space time points are
expected to be effectively smeared to a size of order Planck scale. That is,
it is meaningless to claim, for instance, that you sit exactly on a light
cone.

Many of the problems with the $\alpha $-vacua are related to the image
charges and their apparent non local nature. This is what we turn to next.

\bigskip

\section{Do image charges break causality?}

\bigskip

Another issue that has been brought up concerns the long distance behavior
of the theory. As we have seen, the $\alpha $-vacua can easily be
constructed by allowing for image charges on the wrong side of de Sitter
space. Greens functions might therefore receive contributions directly from
a source but also through the image. This has caused concerns that the
theory does not make physical sense and one might worry about problems of,
e.g., causality.

It is, however, important to bear in mind the physical interpretations of
the various Greens functions. The ones which are important for causality are
the commutator and the retarded Greens functions, which are independent of
the choice of vacuum and always vanish outside of the light cone. When
physics depending on these Greens functions is studied, there is no
difference between the standard Bunch-Davies vacuum and the more general $%
\alpha $-vacua. Greens functions that do depend on the choice of vacua, and
therefore are different for the $\alpha $-vacua, are, e.g., the Feynman
propagator and the Hadamard function. These are the ones that exhibit the
extra singularities, outside of the light cone, that have caused some
worries. One should note, however, that these Greens functions in general
are expected to be non zero outside the light cone; the Bunch-Davies vacuum
is no exception. Contrary to the commutator and retarded Greens function
they encode information about the vacuum and correlations in the vacuum
fluctuations. Unorthodox behavior, like in the case of the $\alpha $-vacua,
might seem surprising but does not necessarily imply that the theory is
inconsistent.

Of particular interest in this context is the Hadamard function. Note that
the Feynman propagator can be written in terms of the vacuum independent
retarded and advanced Greens functions and the vacuum dependent Hadamard
function. The extra singularities of the Hadamard function occur, if we
consider points at equal time in Robertson-Walker coordinates, for points
separated by the diameter of de Sitter space. They simply imply that the
correlation between fluctuations at such points is enhanced. The vanishing
of the commutator guarantees that this can not be used for communications
and can not lead to any break of causality. The large separation between the
points makes it a little difficult for an observer to actually measure the
correlations, although it is suggested in \cite{Einhorn:2002nu} that
interactions might facilitate this. As far as the transplanckian effects are
concerned, this is, however, not really the main point. When inflation ends
the full structure of the Hadamard function will become visible and
available to the CMBR. As suggested in \cite{Danielsson:2002qh}, the end of
inflation transforms the meta observables of \ \cite{Witten:2001kn} into
real observables. Unfortunately, the direct detection of correlations over
distances of the order the de Sitter diameter is not possible. At the end of
inflation, the modes relevant for the CMBR are much larger than the de
Sitter diameter, and the direct as well as the image mediated fluctuations
are hidden in the extremely small scales. The only aspect that is claimed to
be of relevance for the CMBR is a possible large scale tail.

\cite{Kaloper:2002cs} reached the same conclusion as in the present work
concerning the irrelevance of correlations over spatial distances in the
Feynman propagator. The authors of \cite{Kaloper:2002cs} continued, however,
by pointing out that even though the singularities are harmless, a
propagator that behaves like in Minkowsky space for small distances does not
suffer from them anyway. Small distances and Minkowsky behavior can
certainly be reached in the present universe well above the Planck scale.
But, as has already been emphasized, the situation is very different during
inflation when the Planck scale might be just a few orders of magnitude
smaller than the inflationary scale. In such a universe Minkowsky behavior
can not be reached before transplanckian physics make quantum field theory
irrelevant. This way of arguing for the Bunch-Davies vacuum therefore lacks
force.

Finally, it should be noted that it is essential for the consistency of the
vacua that we restrict ourselves to half of de Sitter space. As pointed out
already in \cite{Mottola:ar}, the direction of time changes on the wrong
side of de Sitter. This also guarantees that there is no retarded propagator
connecting an image charge with anything on the right half of de Sitter
space.\footnote{%
Note however recent work, \cite{Parikh:2002py}, where other possibilities to
handle this problem are discussed.}

\bigskip

\section{Can $\protect\alpha $ change with $H$?}

\bigskip

In \cite{Kaloper:2002cs} it is argued that $\alpha $ can not vary with a
changing $H$. The reason, according to \cite{Kaloper:2002cs}, is that local
physics can not know about how $H$ changes and consequently $\alpha $ must
remain constant. Since present day physics requires $e^{\alpha }$ to be
substantially lower than any value that would be of interest to inflation,
and $\alpha $ can not change with time, all interesting effects due to
transplanckian physics are excluded.

However, one should note that even the Bunch-Davies modes have a dependence
on $H$. This is easily seen from the Fourier transform of a Bunch-Davies
mode given by 
\[
\phi _{\mathbf{k,}BD}\left( x\right) =-\eta H\frac{1}{\sqrt{2k}}e^{-ik\eta
}\left( 1-\frac{i}{k\eta }\right) e^{i\mathbf{k}\cdot \mathbf{x}}, 
\]
where $k$ is the comoving momentum. The form when the mode is created at the
fundamental scale is obtained by using $p=k/a=\Lambda $ and recalling that $%
\eta =-\frac{1}{aH}$. The result has a dependence on $H$, and the planckian
physics preparing the modes in a Bunch-Davies state therefore needs to know
about the value of $H$. In a pure quantum field theory world the dependence
is naturally generated through the expansion of the universe while the mode
is in the transplanckian regime. A further dependence on $H$ through $\alpha 
$ is not qualitatively much different and it is easy to see that such a
dependence is quite natural and local. The scale factor depends on time
through $a\left( t\right) =e^{Ht}$, with the expansion of space as a local
effect taking place everywhere. The conjugate time interval associated with
a mode of energy $\Lambda $ is $1/\Lambda $, and during this time interval
space expands by a factor $e^{H\left( t+1/\Lambda \right) -Ht}=e^{H/\Lambda
}\sim 1+\frac{H}{\Lambda }$. This is true also for spatial distances close
to the Planck scale and it is therefore natural to expect effects of the
order $\frac{H}{\Lambda }$. This is immediately obvious from looking at the
Bunch-Davies modes above, but transplanckian physics might lead to an
additional dependence on\ $\frac{H}{\Lambda }$ through $\alpha $.

The simplest, and most natural dependence, is that we have 
\[
e^{\alpha }\sim \frac{H}{\Lambda }, 
\]
and that this remains true at all times. Not only during inflation. However,
in \cite{Starobinsky:2002rp} it has been pointed out that even this small
value of $e^{\alpha }$ could be in conflict with present day measurements of
high energy gamma rays. A vacuum of the type above, it is claimed, would
lead to a higher rate of particle creation than what is acceptable. The
discrepancy is not large, considering the sensitivity of the argument to the
detailed astro-physics involved -- it is a matter of an order of magnitude
or two -- but it nevertheless suggests that a more complicated behavior than
the one above might be needed. This is precisely the subject of \cite
{Goldstein:2002fc} where the vacuum relaxes towards the Bunch-Davies vacuum
once inflation is over. At any rate, high energy gamma rays could, as
suggested in \cite{Starobinsky:2002rp}, be another way to probe planckian
physics in an expanding universe. One should also investigate how these
effects might vary due to local variations in $H$.

\bigskip

\section{Will the $\protect\alpha $-vacua thermalize?}

\bigskip

As discussed in \cite{Bousso:2001mw}\cite{Einhorn:2002nu}\cite
{Kaloper:2002cs} the $\alpha $-vacua are not thermal. One effect of this is
that a detector in an $\alpha $-vacuum will not end up in thermal
equilibrium. An equilibrium will be reached \cite{Einhorn:2002nu}\cite
{Kaloper:2002cs}, but it will not be thermal in character and in general not
obey the laws of detailed balance. For a detector with just two levels,
equilibrium trivially implies detailed balance between the two levels even
though the occupation numbers will be non standard. With three levels,
however, the situation is more unusual. Instead of detailed balance between
any two levels, there will be a net transition rate from, say, level one to
level two, from level two to level three and then back to level one. This is
a consequence of the non thermal nature of the background and does not imply
any inconsistent physics of the detector. A question one can ask, however,
is whether and how such a background can sustain itself. Will there be
processes that act to thermalize the background?

In \cite{Kaloper:2002cs} it is argued based on holography and
complementarity that this will indeed be the case. The main argument is
that, according to a specific observer in de Sitter space, any perturbation
will be redshifted as it approaches the de Sitter horizon and apparently
boiled to pieces through the ever higher temperature that the perturbation
experiences according to the observer. All perturbations will inevitably be
thermalized. One would conclude from this that all traces of physics taking
place on smaller scales than the horizon will be erased and the featureless
and thermal Bunch-Davies modes are the only thing that remains. However,
this is slightly problematic if we take a point of view based on the
Robertson-Walker coordinates in (\ref{metric}). We can now follow the modes
through the de Sitter horizon and see how they freeze when they expand and
become larger than the de Sitter radius. The analogue would be to follow an
observer who ventures inside a black hole and find that nothing peculiar or
dramatic happens near the horizon. And this seems to be the appropriate
point of view to take when discussing fluctuations generated through
inflation. Holography and complementarity are intriguing concepts that could
be important for cosmology, but for this particular exercise they do not
seem to be relevant. As observers of the CMBR we are more in the position of
an observer inside a black hole rather than one on the outside.

A more conservative approach would be to predict the evolution of the
fluctuations without thinking about holography. Any effect that would
threaten to thermalize the $\alpha $-vacua must then be due to interactions.
To be specific one could imagine a thermal background with a small
perturbation that starts out at Planck scale. It is important to note that
the temperature is very low with the thermal wave length of the order of the
de Sitter radius itself. As a consequence a planckian mode is relatively
safe in the beginning. As it redshifts, and its energy decreases towards the
de Sitter temperature, the thermalization effects should become more
important. This happens over a period of time a few times the inverse
temperature. But at the same time the mode will expand past the horizon and
freeze. One seems to conclude that the thermalization effects could be
important in an interacting theory, but it is not clear that they will
suppress any difference with respect to the Bunch-Davies vacuum by orders of
magnitude. Much less wipe out any difference completely. Clearly more
detailed work is needed to make a precise prediction.

\bigskip

\section{Conclusions}

\bigskip

In this short note we have discussed a few of the problems of transplanckian
physics that can be addressed in the simplified framework of $\alpha $%
-vacua. Our conclusion is that the peculiar image singularities of the
Feynman propagator, as well as problems with non renormalization, do not
necessarily imply the ruling out of transplanckian effects. A true
inconsistency at this level would have had dramatic implications. Without
observational input we would have been able to exclude all models of
transplanckian physics that result in other vacua than the Bunch-Davies,
including the non commutative examples discussed in, e.g., \cite
{Easther:2001fi}. We do not believe that such strong conclusions can be
drawn from work done so far. We have furthermore considered the way $\alpha $
might change. We do not see any fundamental difficulties in this respect but
note that the work of \cite{Starobinsky:2002rp} shows that there are
important constraints coming from present day physics. Issues of
thermalization also need further investigation but we do not think that any
argument based on holography put forward so far gives any meaningful
constraints.

To summarize, we believe that the essence of the arguments of \cite
{Kaloper:2002cs} is the following claim: \textit{it is not consistent to
assume a fixed high energy cut off where unknown transplanckian physics
delivers states different form the Bunch-Davies vacuum. }An important point
of the present paper is that possible problems are not visible in the cutoff
theory. All problems must rely on claims about contributions from energy
scales above the cutoff, energy densities etc., and speculations about
transplanckian physics. Clearly, more work is needed to figure out how
unorthodox transplanckian physics really can be.

\section*{Acknowledgments}

The author would like to thank Lars Bergstr\"{o}m, Daniel Domert, Martin
Olsson, Hector Rubinstein, Gary Shiu and Konstantin Zarembo for valuable
discussions. The author is a Royal Swedish Academy of Sciences Research
Fellow supported by a grant from the Knut and Alice Wallenberg Foundation.
The work was also supported by the Swedish Research Council (VR).


\bigskip

\begin{thebibliography}{99}
\bibitem{Brandenberger:1999sw}  R.~H.~Brandenberger, ``Inflationary
cosmology: Progress and problems,'' arXiv:hep-ph/9910410. 


\bibitem{Martin:2000xs}  J.~Martin and R.~H.~Brandenberger, ``The
trans-Planckian problem of inflationary cosmology,'' Phys.\ Rev.\ D \textbf{%
63}, 123501 (2001) [arXiv:hep-th/0005209]. 


\bibitem{Niemeyer:2000eh}  J.~C.~Niemeyer, ``Inflation with a high frequency
cutoff,'' Phys.\ Rev.\ D \textbf{63}, 123502 (2001)
[arXiv:astro-ph/0005533]. 


\bibitem{Brandenberger:2000wr}  R.~H.~Brandenberger and J.~Martin, ``The
robustness of inflation to changes in super-Planck-scale physics,'' Mod.\
Phys.\ Lett.\ A \textbf{16}, 999 (2001) [arXiv:astro-ph/0005432]. 


\bibitem{Kempf:2000ac}  A.~Kempf, ``Mode generating mechanism in inflation
with cutoff,'' Phys.\ Rev.\ D \textbf{63}, 083514 (2001)
[arXiv:astro-ph/0009209]. 


\bibitem{Chu:2000ww}  C.~S.~Chu, B.~R.~Greene and G.~Shiu, ``Remarks on
inflation and noncommutative geometry,'' Mod.\ Phys.\ Lett.\ A \textbf{16},
2231 (2001) [arXiv:hep-th/0011241]. 


\bibitem{Martin:2000bv}  J.~Martin and R.~H.~Brandenberger, ``A cosmological
window on trans-Planckian physics,'' arXiv:astro-ph/0012031. 

\bibitem{Mersini:2001su}  L.~Mersini, M.~Bastero-Gil and P.~Kanti, ``Relic
dark energy from trans-Planckian regime,'' Phys.\ Rev.\ D \textbf{64},
043508 (2001) [arXiv:hep-ph/0101210]. 

\bibitem{Niemeyer:2001qe}  J.~C.~Niemeyer and R.~Parentani,
``Trans-Planckian dispersion and scale-invariance of inflationary
perturbations,'' Phys.\ Rev.\ D \textbf{64}, 101301 (2001)
[arXiv:astro-ph/0101451]. 


\bibitem{Kempf:2001fa}  A.~Kempf and J.~C.~Niemeyer, ``Perturbation spectrum
in inflation with cutoff,'' Phys.\ Rev.\ D \textbf{64}, 103501 (2001)
[arXiv:astro-ph/0103225]. 

\bibitem{Starobinsky:2001kn}  A.~A.~Starobinsky, ``Robustness of the
inflationary perturbation spectrum to trans-Planckian physics,'' Pisma Zh.\
Eksp.\ Teor.\ Fiz.\ \textbf{73}, 415 (2001) [JETP Lett.\ \textbf{73}, 371
(2001)] [arXiv:astro-ph/0104043].

\bibitem{Easther:2001fi}  R.~Easther, B.~R.~Greene, W.~H.~Kinney and
G.~Shiu, ``Inflation as a probe of short distance physics,'' Phys.\ Rev.\ D 
\textbf{64}, 103502 (2001) [arXiv:hep-th/0104102]. 

\bibitem{Bastero-Gil:2001rv}  M.~Bastero-Gil and L.~Mersini, ``SN1A data and
CMB of Modified Curvature at Short and Large Distances,'' Phys.\ Rev.\ D 
\textbf{65} (2002) 023502 [arXiv:astro-ph/0107256]. 

\bibitem{Hui:2001ce}  L.~Hui and W.~H.~Kinney, ``Short distance physics and
the consistency relation for scalar and tensor fluctuations in the
inflationary universe,'' arXiv:astro-ph/0109107. 

\bibitem{Easther:2001fz}  R.~Easther, B.~R.~Greene, W.~H.~Kinney and
G.~Shiu, ``Imprints of short distance physics on inflationary cosmology,''
arXiv:hep-th/0110226. 

\bibitem{Bastero-Gil:2001nu}  M.~Bastero-Gil, P.~H.~Frampton and L.~Mersini,
``Modified dispersion relations from closed strings in toroidal cosmology,''
arXiv:hep-th/0110167. 

\bibitem{Brandenberger:2001ty}  R.~H.~Brandenberger, S.~E.~Joras and
J.~Martin, ``Trans-Planckian physics and the spectrum of fluctuations in a
bouncing universe,'' arXiv:hep-th/0112122. 

\bibitem{Martin:2002kt}  J.~Martin and R.~H.~Brandenberger, ``The
Corley-Jacobson dispersion relation and trans-Planckian inflation,''
arXiv:hep-th/0201189. 

\bibitem{Niemeyer:2002ze}  J.~C.~Niemeyer, ``Cosmological consequences of
short distance physics,'' arXiv:astro-ph/0201511. 

\bibitem{Lizzi:2002ib}  F.~Lizzi, G.~Mangano, G.~Miele and M.~Peloso,
``Cosmological perturbations and short distance physics from noncommutative
geometry,'' arXiv:hep-th/0203099. 

\bibitem{Shiu:2002kg}  G.~Shiu and I.~Wasserman, ``On the signature of short
distance scale in the cosmic microwave background,'' arXiv:hep-th/0203113.

\bibitem{Brandenberger:2002nq}  R.~Brandenberger and P.~M.~Ho,
``Noncommutative spacetime, stringy spacetime uncertainty principle, and
density fluctuations,'' arXiv:hep-th/0203119. 

\bibitem{Shankaranarayanan:2002ax}  S.~Shankaranarayanan, ``Is there an
imprint of Planck scale physics on inflationary cosmology?,''
arXiv:gr-qc/0203060. 

\bibitem{Kaloper:2002uj}  N.~Kaloper, M.~Kleban, A.~E.~Lawrence and
S.~Shenker, ``Signatures of short distance physics in the cosmic microwave
background,'' arXiv:hep-th/0201158. 

\bibitem{Brandenberger:2002hs}  R.~H.~Brandenberger and J.~Martin, ``On
signatures of short distance physics in the cosmic microwave background,''
arXiv:hep-th/0202142. 

\bibitem{Hassan:2002qk}  S.~F.~Hassan and M.~S.~Sloth, ``Trans-Planckian
effects in inflationary cosmology and the modified uncertainty principle,''
arXiv:hep-th/0204110.

\bibitem{Danielsson:2002kx}  U.~H.~Danielsson, ``A note on inflation and
transplanckian physics,'' Phys.\ Rev.\ D \textbf{66}, 023511 (2002)
[arXiv:hep-th/0203198]. 

\bibitem{Easther:2002xe}  R.~Easther, B.~R.~Greene, W.~H.~Kinney and
G.~Shiu, ``A generic estimate of trans-Planckian modifications to the
primordial power spectrum in inflation,'' arXiv:hep-th/0204129.

\bibitem{Danielsson:2002qh}  U.~H.~Danielsson, ``Inflation, holography and
the choice of vacuum in de Sitter space,'' JHEP \textbf{0207}, 040 (2002)
[arXiv:hep-th/0205227]. 

\bibitem{Niemeyer:2002kh}  J.~C.~Niemeyer, R.~Parentani and D.~Campo,
``Minimal modifications of the primordial power spectrum from an adiabatic
short distance cutoff,'' arXiv:hep-th/0206149. 

\bibitem{chernikov}  N. A. Chernikov and E. A. Tagirov, ``Quantum theory of
scalar field in de Sitter space-time,'' Ann. Inst. Henri Poincar\'{e}, vol.
IX, nr 2, (1968) 109.

\bibitem{Mottola:ar}  E.~Mottola, ``Particle Creation In De Sitter Space,''
Phys.\ Rev.\ D \textbf{31} (1985) 754.

\bibitem{Allen:ux}  B.~Allen, ``Vacuum States In De Sitter Space,'' Phys.\
Rev.\ D \textbf{32} (1985) 3136.

\bibitem{Floreanini:1986tq}  R.~Floreanini, C.~T.~Hill and R.~Jackiw,
``Functional Representation For The Isometries Of De Sitter Space,'' Annals
Phys.\ \textbf{175} (1987) 345.

\bibitem{Bousso:2001mw}  R.~Bousso, A.~Maloney and A.~Strominger,
``Conformal vacua and entropy in de Sitter space,'' arXiv:hep-th/0112218.

\bibitem{Spradlin:2001nb}  M.~Spradlin and A.~Volovich, ``Vacuum states and
the S-matrix in dS/CFT,'' arXiv:hep-th/0112223.

\bibitem{Banks:2002nv}  T.~Banks and L.~Mannelli, ``De Sitter vacua,
renormalization and locality,'' arXiv:hep-th/0209113. 

\bibitem{Einhorn:2002nu}  M.~B.~Einhorn and F.~Larsen, ``Interacting Quantum
Field Theory in de Sitter Vacua,'' arXiv:hep-th/0209159. 

\bibitem{Kaloper:2002cs}  N.~Kaloper, M.~Kleban, A.~Lawrence, S.~Shenker and
L.~Susskind, 
arXiv:hep-th/0209231.

\bibitem{Parikh:2002py}  M.~K.~Parikh, I.~Savonije and E.~Verlinde,
``Elliptic de Sitter space: dS/Z(2),'' arXiv:hep-th/0209120. 

\bibitem{Witten:2001kn}  E.~Witten, ``Quantum gravity in de Sitter space,''
arXiv:hep-th/0106109. 

\bibitem{Starobinsky:2002rp}  A.~A.~Starobinsky and I.~I.~Tkachev,
``Trans-Planckian particle creation in cosmology and ultra-high energy
cosmic rays,'' arXiv:astro-ph/0207572. 

\bibitem{Goldstein:2002fc}  K.~Goldstein and D.~A.~Lowe, 
arXiv:hep-th/0208167. 
\end{thebibliography}
\end{document}